\documentclass[aps,prl,a4paper,preprint]{revtex4}
\usepackage{amsfonts,amssymb}
\usepackage{graphicx}

\begin{document}

\draft
\title{THEORY OF SPIN-WAVE FREQUENCY GAPS
IN 3D MAGNONIC CRYSTALS.  APPLICATION TO MANGANITES.}
\author{M. Krawczyk, H. Puszkarski}
\address{Surface Physics Division, Faculty of Physics, Adam
Mickiewicz University,  ul. Umultowska 85, Pozna\'{n}, 61-614 Poland.}

\date{\today}

\begin{abstract}
This study is an investigation of spin wave spectrum in macrostructures composed of two ferromagnetic materials and showing a 3D periodicity:
spherical ferromagnetic grains disposed in the nodes of a 3D crystal lattice are embedded in a matrix with different ferromagnetic properties.
Frequency ranges forbidden to spin wave propagation are found in the calculated magnonic spectra. Both the position and the width of the gaps
are found to depend on the magnetic (exchange and magnetization) contrasts in the composite material, as well as on its structural parameters
(filling fraction, crystal lattice type). Having applied our theory to interpretation of the existence of a spin wave gap in doped manganites,
recently revealed in neutron scattering experiments by S. Hennion {\em et al.}, we obtained a good (though approximate) quantitative agreement
with the experimental results. A working hypothesis is proposed on this basis, supposing that certain manganites can be treated as natural 3D
magnonic crystals.
\end{abstract}
\maketitle

\section{Introduction}

Though the first study of electromagnetic wave propagation in
periodic structures, written by lord Rayleigh, was published
already in 1887, it is only in recent years that photonic
composites raised suddenly an extremely keen interest. The
research in this field was initiated by the studies by
Yablonovitch and John \cite{[1],[2]} published in 1987 and
anticipating the existence of complete energy gaps in
electromagnetic wave spectra in three-dimensional periodic
composites, henceforth referred to as {\em photonic crystals}.
These can be used for fabricating new optoelectronic devices in
which the role of electrons, traditionally used as transport
medium, would be played by photons \cite{[3],[4],[5]}. The
so-called {\em left-handed materials} (LHM), showing negative
effective refractivity \cite{[6],[7]}, provide an example of
periodic materials demonstrating how much the properties of this
kind of structure can differ from those of homogeneous materials.
Another type of periodic composites are structures composed of
materials with different {\em elastic} properties; showing an
energy gap in their {\em elastic} wave spectrum, such composites
are referred to as {\it phononic crystals} \cite{[8]}-\cite{[10]}.
Recently, attention has been focused on the search of photonic and
phononic crystals in which both the position and the width of the
energy gap could be controlled by external factors, such as
applied voltage or magnetic field. Attempts are made to create
photonic crystals in which one of the component materials would be
a magnetic \cite{[11]}-\cite{[15]}.

A magnetic periodic composite consists of at least two magnetic
materials; the information carrier in such structures are spin
waves. By analogy to photonic and phononic crystals, in which the
role of information carrier is played by photons and phonons
respectively, periodic magnetic composites are referred to as {\em
magnonic crystals}. Studies of 2D magnonic crystals have been
reported \cite{[16]}-\cite{[21]}, with scattering centers in the
form of "infinitely" long cylinders disposed in square lattice
nodes (cylinder and matrix materials being two different
ferromagnetics), and the anticipated gaps found indeed in the
respective spin wave spectra. Further research was focused on
magnetic multilayer systems, which can be regarded as 1D magnonic
crystals \cite{[22]}-\cite{[30]}.

In this paper, we present numerically calculated magnonic band
structures - to our best knowledge not yet reported in the
literature - of {\em three-dimensional} magnonic crystals. Due to
the complexity of the problem, only the simplest model of 3D
magnonic crystal is considered here, represented by a system of
ferromagnetic spheres (which act as scattering centers) disposed
in the nodes of a cubic crystal lattice embedded in a different
ferromagnetic material (matrix). Both the exchange and dipolar
interactions are taken into account in our calculations, based on
the plane wave method and using the linear approximation. As a
conclusion, we propose a new magnonic interpretation of
experimental results obtained through neutron scattering on spin
waves in doped manganites \cite{[32]}-\cite{[35]}.

\section{The procedure of numerical calculation of 3D magnonic band structure}

\begin{figure}[h]
\begin{center}
\includegraphics[width=90mm]{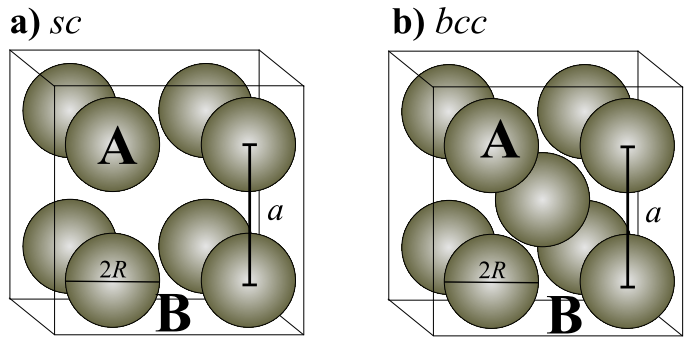}


\caption{The 3D periodic structures studied in this paper; two
types of structure are considered, both consisting of
ferromagnetic spheres of material {\bf A} embedded in a matrix of
material {\bf B} (materials {\bf A} and {\bf B} have different
magnetic properties); (a) the spheres are disposed in the nodes of
an {\em sc} lattice; (b) the spheres are disposed in the nodes of
a {\em bcc} lattice. \label{fig1}}
\end{center} \end{figure}

Let's consider an ideal periodic structure consisting of spheres
of ferromagnetic {\bf A} and embedded in a matrix of ferromagnetic
{\bf B}. The spheres are assumed to form a 3D periodical lattice
of $sc$ (Fig. \ref{fig1}a) or {\em bcc} (Fig. \ref{fig1}b) type. A
static magnetic field, $H_{0}$, is applied to the composite along
the $x_{3}$ axis, and assumed to be strong enough to saturate the
magnetization of both materials. The lattice constant is denoted
by $a$; the filling fraction, $f$, is defined as the volume
proportion of material {\bf A} in a unit cell:
\begin{equation}
f={\displaystyle \frac{\frac{4}{3}\pi R^{3}}{a^{3}}} \;\;\;
\mbox{({\em sc})}\;\;\; \;\;\mbox{or}\;\;\;\;\;
f={\displaystyle\frac{\frac{8}{3}\pi R^{3}}{a^{3}}}\;\;\;
\mbox{({\em bcc}),} \label{eq1} \end{equation} in an {\em sc} or
{\em bcc} lattice, respectively.

Ferromagnetics {\bf A} and {\bf B} are characterized by two
material parameters: the spontaneous magnetization ($M_{S,A}$ and
$M_{S,B}$), and the exchange constant ($A_{A}$ and $A_{B}$); both
these parameters depend on the position vector
$\vec{r}=(x_{1},x_{2},x_{3})$:
\begin{eqnarray}
M_{S}(\vec{r})&=&M_{S,B}+(M_{S,A}-M_{S,B})S(\vec{r}),
\nonumber \\
A(\vec{r})&=&A_{B}+(A_{A}-A_{B})S(\vec{r}),\label{eq2}
\end{eqnarray}
the value of function $S(\vec{r})$ being 1 inside a sphere, and 0
beyond.

In the classical approximation, spin waves are described by the
Landau-Lifshitz (LL) equation, taking the following form in the
case of magnetic composites (with damping neglected):
\begin{eqnarray}
\frac{\partial \vec{M}(\vec{r},t)}{\partial t} = \gamma \mu_{0}
\vec{M}(\vec{r},t) \times \vec{H}_{eff}(\vec{r},t), \label{eq3}
\end{eqnarray}
where magnetization $\vec{M}(\vec{r},t)$ is a function of position
vector $\vec{r}$ and time $t$;\\ $\vec{H}_{eff}(\vec{r},t)$ stands
for the effective magnetic field acting on magnetization
$\vec{M}(\vec{r},t)$ and defined as follows
\cite{[17]}-\cite{[19]}:
\begin{eqnarray}
\vec{H}_{eff}(\vec{r},t)=H_{0}\vec{e}_{3}
+\vec{h}(\vec{r},t)+\frac{2}{\mu_{0}} \left(\nabla \cdot
\frac{A}{M^{2}_{S}}\nabla \right) \vec{M}(\vec{r},t); \label{eq4}
\end{eqnarray}
$\vec{e}_{i}$ is the unit vector along the $x_{i}$ axis;
$\vec{h}(\vec{r}, t)$ is the dynamic magnetic field due to dipolar
interactions; the third component represents the exchange field.
The magnetization vector can be represented as the sum of its
static component, $M_{S}\vec{e}_{3}$, which is parallel to the
applied field, and its dynamic component, $\vec{m}(\vec{r},t)$,
lying in plane ($x_{1},x_{2}$):
\begin{eqnarray}
\vec{M}(\vec{r},t)=M_{S}\vec{e}_{3}+\vec{m}(\vec{r},t).
\label{eq5}
\end{eqnarray}
The dynamic dipolar field, $\vec{h}$, must satisfy the
magnetostatic Maxwell equations:
\begin{eqnarray}
\nabla \times \vec{h}(\vec{r})=0, \nonumber \\
\nabla \cdot \left(\vec{h}(\vec{r})+ \vec{m}(\vec{r})\right)=0.
\label{eq6}
\end{eqnarray}
In magnonic crystals, the position-dependent coefficients in
(\ref{eq3}), {\em i.e.} $M_{S}$ and $A$, are periodic functions of
the position vector, which allows us to use, in the procedure of
solving the LL equation defined in (\ref{eq3}), the plane wave
method, described in detail in our earlier papers \cite{[17],[18]}
(dealing with {\em two-dimensional} magnonic crystals). Following
this scheme, we proceed to Fourier-expanding all the periodic
functions of the position vector, {\em i.e.} the spontaneous
magnetization, $M_{S}$, and parameter $Q$ defined as follows:
\begin{eqnarray}
Q=\frac{2A}{\mu_{0}M^{2}_{S}H_{0}}. \label{eq7}
\end{eqnarray}

The dynamic components of the magnetization can be expressed as
the product of the periodic envelope function and the Bloch factor
$exp(i\vec{K}\vec{r}$) ($\vec{K}$ denoting a 3D wave vector); the
envelope function can be transformed into the reciprocal space as
well. Including all the expansions into (\ref{eq4}) and
(\ref{eq6}) leads to the following infinite system of linear
equations for Fourier coefficients of the dynamic magnetization
components, $\vec{m}_{1 \vec{k}}(\vec{g})$  and $\vec{m}_{2
\vec{k}}(\vec{g})$:
\begin{eqnarray}
i \Omega m_{1 \vec{k}}(\vec{g})&=& m_{2 \vec{k}}(\vec{g}) +
\frac{1}{H_{0}}\sum_{\vec{g}'} \frac{(k_{2} +
g'_{2})(k_{1}+g'_{1})}{| \vec{k}+
 \vec{g}' |^{2}} m_{1 \vec{k}}(\vec{g}') M_{S}(\frac{2 \pi}{a}(\vec{g}
- \vec{g}'))\nonumber \\
& + &\frac{1}{H_{0}} \sum_{\vec{g}'} \frac{(k_{2} + g'_{2})^{2}}
{| \vec{k} + \vec{g}' |^{2}} m_{2 \vec{k}}(\vec{g}')
M_{S}(\frac{2 \pi}{a}(\vec{g} - \vec{g}')) \nonumber \\
& +& \sum_{\vec{g}'} \sum_{\vec{g}''} \left(\frac{2
\pi}{a}\right)^{2}[ (\vec{k} + \vec{g}') \cdot (\vec{k}+\vec{g}'')
-
 (\vec{g}-\vec{g}'') \cdot (\vec{g} - \vec{g}' ) ]\nonumber \\
& \cdot & M_{S}(\frac{2 \pi}{a}(\vec{g}-\vec{g}'')) Q(\frac{2
\pi}{a}(\vec{g}'' - \vec{g}')) m_{2 \vec{k}} (\vec{g}'),\nonumber
\end{eqnarray}
\begin{eqnarray}
i \Omega m_{2 \vec{k}}(\vec{g})&=&-m_{1 \vec{k}} (\vec{g})+
-\frac{1}{H_{0}}\sum_{\vec{g}'} \frac{(k_{2} +
 g'_{2})(k_{1}+g'_{1})}{| \vec{k} +
 \vec{g}' |^{2}} m_{2 \vec{k}}(\vec{g}') M_{S}(\frac{2 \pi}{a}(\vec{g}
- \vec{g}'))\nonumber \\
&-& \frac{1}{H_{0}} \sum_{\vec{g}'} \frac{(k_{1} + g'_{1})^{2}} {|
\vec{k} + \vec{g}' |^{2}} m_{1 \vec{k}}(\vec{g}')
M_{S}(\frac{2 \pi}{a}(\vec{g} - \vec{g}'))\nonumber \\
&-&\sum_{\vec{g}'} \sum_{\vec{g}''} \left(\frac{2
\pi}{a}\right)^{2}[ (\vec{k} + \vec{g}') \cdot (\vec{k}+\vec{g}'')
-
 (\vec{g}-\vec{g}'') \cdot (\vec{g} - \vec{g}' ) ] \nonumber \\
&\cdot & M_{S}(\frac{2 \pi}{a}(\vec{g}-\vec{g}'')) Q(\frac{2
\pi}{a}(\vec{g}'' - \vec{g}')) m_{1 \vec{k}}
(\vec{g}').\label{eq8}
\end{eqnarray}
$\vec{k}_{i}$ and $\vec{g}_{i}$ denoting the Cartesian components
of a dimensionless wave vector $\vec{k} = (a/2\pi)\vec{K}$ and a
dimensionless reciprocal lattice vector $\vec{g} =
(a/2\pi)\vec{G}$, respectively; a new quantity introduced in
(\ref{eq8}) is $\Omega$, henceforth referred to as {\em reduced
frequency}:
\begin{eqnarray}
\Omega=\frac{\omega}{|\gamma| \mu_{0} H_{0}}. \label{eq9}
\end{eqnarray}

The Fourier coefficients of spontaneous magnetization $M_{S}$ and
parameter $Q$ are calculated from the inverse Fourier
transformation; in the case of spheres the resulting formulae read
as follows:
\begin{eqnarray}
M_{S}(\vec{G}) = \left\{ \begin{array}{ll}
     M_{S,A}f + M_{S,B}(1-f),& \; \;
     \mbox{for  $\;\vec{G}=0$} \\ & \\
f  (M_{S,A}-M_{S,B})
{\displaystyle\frac{3\left[\sin(GR)-(GR)\cos(GR)\right]}{(GR)^{3}}},&
\; \; \mbox{for $\;\vec{G} \neq 0$}
    \end{array}
    \right., \label{aa2}
\end{eqnarray}
and similarly
\begin{eqnarray}
Q(\vec{G}) = \left\{ \begin{array}{ll}
     Q_{A}f + Q_{B}(1-f),& \; \; \mbox{for  $\;\vec{G}=0$} \\
        & \\
f  (Q_{A}-Q_{B}){\displaystyle \frac{
3\left[\sin{(GR)}-(GR)\cos(GR)\right]}{(GR)^{3}}},& \; \;
\mbox{for $\;\vec{G} \neq 0$}
    \end{array}
    \right., \label{aa1}
\end{eqnarray}
where $Q_{A}$ and $Q_{B}$ are the values of parameter $Q$ in
material {\bf A} and {\bf B}, respectively; $R$ is the sphere
radius (Fig. \ref{fig1}).

Obviously, the numerical calculations performed on the basis of
(\ref{eq8}) involve a finite number of reciprocal lattice vectors
$\vec{G}$ in the Fourier expansions; however, we have made sure
that the number used is large enough to ensure good convergence of
the numerical results. As indicated by an analysis performed, a
satisfactory convergence is obtained already with 343 reciprocal
lattice vectors used.
\begin{figure}[h]
\begin{center}
\includegraphics[width=100mm]{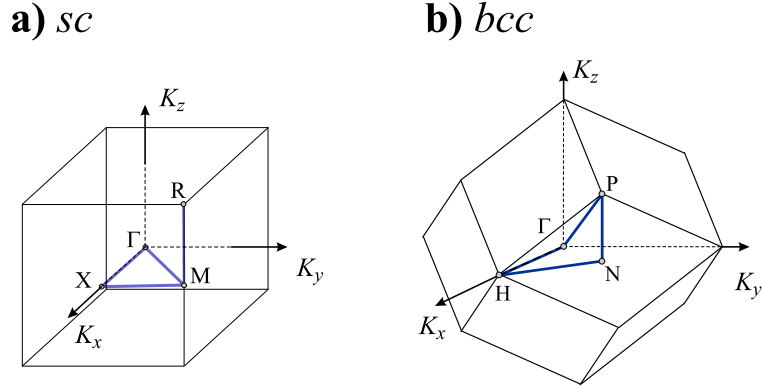}


\caption{. The Brillouin zones corresponding to the considered
structures, and the paths (highlighted) along which the magnonic
band structures are calculated: (a) $M \Gamma XMR$ in the case of
the $sc$ lattice, (b) $P \Gamma HNP$ in the case of the $bcc$
lattice. \label{fig2}}
\end{center} \end{figure}

\section{Numerical results}

\begin{figure}[h]
\begin{center}
\includegraphics[width=55mm]{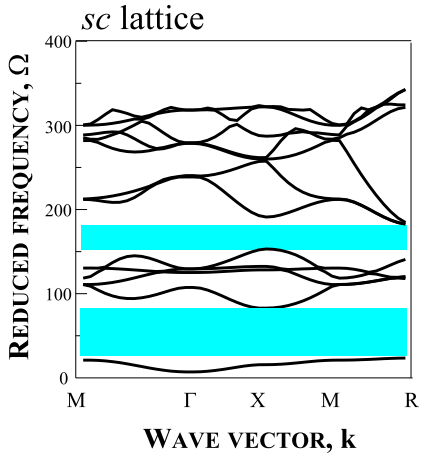}


\caption{The magnonic band structure found numerically for the
$sc$ lattice-based composite (iron spheres embedded in a YIG
matrix). The spin wave energy branches have been calculated along
the path (in the first Brillouin zone) shown in Fig. \ref{fig2}a.
The following parameter values are assumed: applied static
magnetic field $\mu_{0}H_{0}=0.1$T, crystal lattice constant
$a=100\AA$, and filling fraction $f$=0.2. \label{fig3}}
\end{center} \end{figure}

The first 3D magnonic crystal to be studied here is a magnetic
composite consisting of ferromagnetic spheres (of material {\bf
A}) disposed in the nodes of a {\em simple cubic} ({\em sc})
lattice and embedded in a different magnetic material ({\bf B}),
referred to as matrix (Fig. \ref{fig1}a). The corresponding
magnonic band structure will be calculated along path
$M=\pi/a(1,1,0) \rightarrow \Gamma= \pi/a(0,0,0) \rightarrow
X=\pi/a(1,0,0) \rightarrow M=\pi/a(1,1,0) \rightarrow
R=\pi/a(1,1,1)$ in the nonreducible part of the first Brillouin
zone (see Fig. \ref{fig2}a). Iron and yttrium iron garnet (YIG)
are chosen as component materials {\bf A} (spheres) and {\bf B}
(matrix), respectively, in the studied example. As established in
our earlier studies \cite{[17],[18]}, in the case of 2D magnonic
crystals, such composition, involving a substantial contrast of
the magnetic parameters between the component materials
($M_{S,Fe}=1.752\; 10^{6} Am^{-1}$, $M_{S,YIG}=0.194 \;
10^{6}Am^{-1}$, $A_{Fe}=2.1 \; 10^{-11}Jm^{-1}$, $A_{YIG}=0.4 \;
10^{-11}Jm^{-1}$), leads to large energy gaps in the spin wave
spectrum. Also in the spin wave spectrum obtained for the sc
lattice-based 3D magnonic crystal considered here, two wide energy
gaps are present, one between bands 1 and 2, the other between
bands 5 and 6, as shown in Fig. \ref{fig3}. The existence of these
gaps signifies that spin waves having frequency values within a
gap cannot propagate in the composite. The following parameter
values have been assumed in our calculations: crystal lattice
constant $a$=100$\AA$, applied static magnetic field
$\mu_{0}H_{0}=0.1$T, and filling fraction $f$=0.2.

\begin{figure}[h]
\begin{center}
\includegraphics[height=50mm]{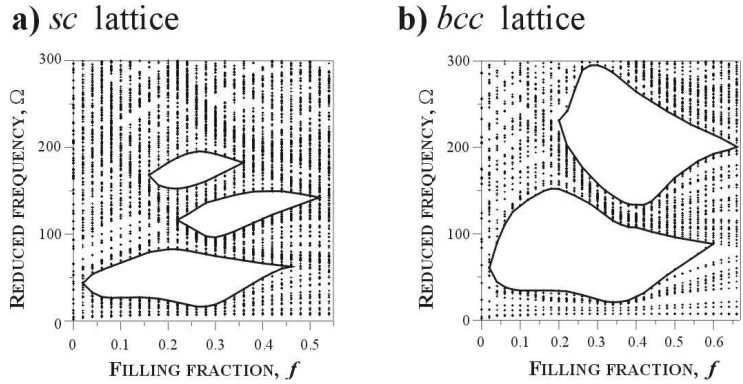}

\caption{The magnonic energy branches in cubic magnonic crystals
($Fe$ spheres embedded in YIG) plotted versus the filling fraction
(a) for an $sc$ lattice-based structure, and (b) for a $bcc$
lattice-based structure. The lattice constant value is assumed to
be $a=100\AA$, and the applied magnetic field value to be
$\mu_{0}H_{0}=0.1$T. \label{fig4}}
\end{center} \end{figure}

Let's examine the effect of the filling fraction on the width of
the gaps obtained. Figure \ref{fig4}a shows the magnonic bands
plotted against the filling fraction in the {\em sc} lattice-based
structure ($Fe$ spheres embedded in YIG); the plot was obtained
through projecting the magnonic band structure calculated along
$M\Gamma XMR$ (for a fixed filling fraction value) on the reduced
frequency ($\Omega$) axis; the procedure of projection was
performed repeatedly for consecutive filling fraction values. The
first (lowest) gap shown in Fig. \ref{fig4}a is found to be the
largest for filling fraction value $f$=0.24, the corresponding gap
width being $\Delta \Omega$=63.29, while the much narrower second
gap reaches its maximum width at $f$=0.3. Unlike in 1D and 2D
magnonic crystals, no oscillatory variations of the energy gaps
with the filling fraction are observed in the 3D composites
\cite{[17],[18],[22]}. The gaps found in the {\em bcc}
lattice-based structure, shown in Fig. \ref{fig4}b, are clearly
much broader than those in the {\em sc} lattice-based structure.
The spin wave reduced frequency was calculated along path $P\Gamma
HNP$ [$P=\pi/a(1,1,1) \rightarrow \Gamma= \pi/a(0,0,0) \rightarrow
H=\pi/a(2,0,0) \rightarrow N=\pi/a(1,1,0) \rightarrow
P=\pi/a(1,1,1)$] (indicated by a bold line in Fig. \ref{fig2}b) in
the {\em bcc} lattice first Brillouin zone. The maximum width of
the first gap is found to occur at $f$=0.2, reaching value $\Delta
\Omega$=117.02, which is 85\% larger than its counterpart in the
{\em sc} lattice-based structure. The second energy gap found in
the {\em bcc} lattice-based structure (between bands 4 and 5) is
broader than the first one, and reaches a maximum width of $\Delta
\Omega$ =141.52 at filling fraction value $f$=0.32.

\begin{figure}[h]
\begin{center}
\includegraphics[height=50mm]{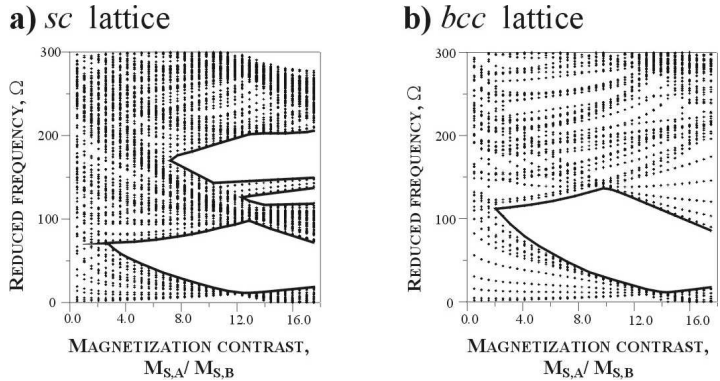}


\caption{Three-dimensional magnonic crystal spin wave spectra
plotted versus the magnetization contrast, on the basis of
computations performed for (a) an $sc$ lattice-based structure,
and (b) a $bcc$ lattice-based structure. The assumed spontaneous
magnetization and exchange constant values in the matrix are
$M_{S,B}=0.194\; 10^{6}Am^{-1}$ and $A_{B}=0.4\; 10^{-11}Jm^{-1}$,
close to the respective parameter values in YIG; the exchange
constant value in the spheres is assumed to be $A_{A}=2.1\;
10^{-11}Jm^{-1}$, the lattice constant $a=100\AA$, the filling
fraction $f=0.2$, and the applied field $\mu_{0}H_{0}=0.1T$.
\label{fig5}}
\end{center} \end{figure}

Let's proceed now to the role played by the magnetic parameters of
the component materials. Figure \ref{fig5} shows spin wave energy
branches plotted against the contrast between the spontaneous
magnetization values in materials {\bf A} (spheres) and {\bf B}
(matrix); this contrast, defined as ratio $M_{S,A}/M_{S,B}$, will
be henceforth shortly referred to as {\em magnetization contrast}.
The computations have been performed for a fictitious matrix
material whose spontaneous magnetization and exchange constant
values, $M_{S,B}$ and $A_{B}$, respectively, are fixed at values
close to those in YIG; the exchange constant value in the sphere
material is assumed to be $A_{A}=2.1\; 10^{-11} Jm^{-1}$. The
other parameters are fixed as well: the assumed lattice constant
value is $a$=100$\AA$, the filling fraction $f$=0.2, and the
applied magnetic field $\mu_{0}H_{0}=0.1$T. The results obtained
for the {\em sc} lattice-based and the {\em bcc} lattice-based
structures are depicted in Figs. 5a and 5b, respectively. Let's
focus on the first gap occurring between the first and the second
energy bands; the gap width is found to be maximal for
magnetization contrast values 10 and 12.8 in the {\em bcc}
lattice-based and the sc lattice-based structures, respectively.
Both these values are not far from the spontaneous magnetization
contrast between iron and YIG ($M_{S,Fe} / M_{S,YIG}$ = 10.82),
the material composition used in the above-discussed investigation
of the effect of the filling fraction, and leading to quite wide
gaps, as one can see in Figs. \ref{fig4}a and \ref{fig4}b. Note
that in both lattice types the minimum spontaneous magnetization
contrast necessary for energy gaps to open is greater than 2.

\begin{figure}[h]
\begin{center}
\includegraphics[height=50mm]{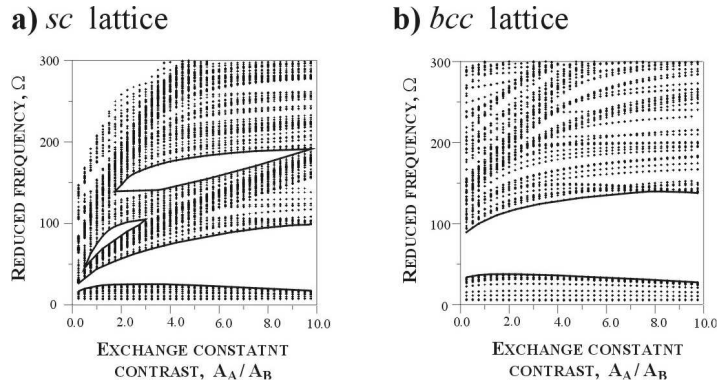}


\caption{Magnonic branches plotted versus the exchange contrast in
the composite. The assumed values of the matrix material
parameters are close to the respective parameter values in YIG;
the spontaneous magnetization in the spheres is assumed to be
$M_{S,A}=2.1\; 10^{6}Am^{-1}$. Other parameter values are as in
Fig. 5. \label{fig6}}
\end{center} \end{figure}

In the last stage of this investigation, we shall examine the role
of the contrast between the exchange constant values in the
component materials. The results of the respective computations
are depicted in Fig. \ref{fig6}, showing magnonic branches {\em
versus} the {\em exchange contrast}, defined as the ratio
($A_{A}/A_{B}$) between the exchange constant values in materials
{\bf A} (spheres) and {\bf B} (matrix); results obtained for {\em
sc} and {\em bcc} lattice-based structures are shown in Figs.
\ref{fig6}a and \ref{fig6}b, respectively. The only variable in
the computations was the exchange constant in material {\bf A},
all the other structural and material parameters being fixed:
$M_{S,A}=1.752\; 10^{6}Am^{-1}$, $M_{S,B}=0.194\; 10^{6}Am^{-1}$,
$A_{B}=0.4\; 10^{-11}Jm^{-1}$, $a=100\AA$, $\mu_{0}H_{0}=0.1$T.
The occurrence of the first energy gap (between the first and the
second magnonic bands) is found to be independent of the exchange
contrast, yet the gap width grows as the exchange contrast
increases. This means that the exchange contrast is not an
indispensable factor for the opening of complete energy gaps in 3D
magnonic crystals, its effect being limited to the width of the
{\em existing} gaps.

\section{Confrontation of the theory with the experiment}

As demonstrated by the above-presented results, complete energy
gaps can occur in the spin wave spectra of 3D periodic magnetic
composites. The opening of such gaps is conditioned on the
material composition, as gaps are found to exist only when the
spontaneous magnetization contrast is greater than 2. The gap
width can be controlled through adjusting the filling fraction
value, and the position of the gap on the energy scale through
modifying the value of the applied magnetic field. As expected,
the {\em bcc} lattice-based structure is found to give rise to
larger energy gaps than the {\em sc} lattice-based one.

\begin{figure}[h]
\begin{center}
\includegraphics[width=55mm]{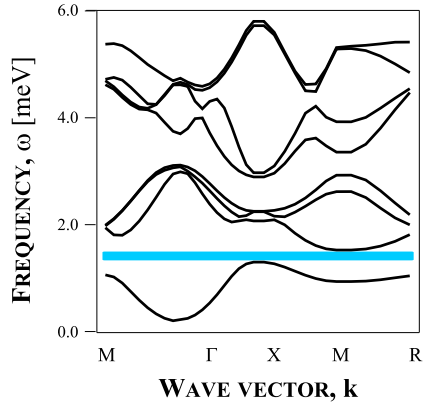}


\caption{Magnonic band structure computed on the basis of Eqs
(\ref{eq8}) for a hypothetic periodic structure of ferromagnetic
droplets existing in doped manganites $La_{1-x}Ca_{x}MnO_{3}$
$(1\%<x<9\%)$; the droplets are assumed to be sphere-shaped and to
form a regular crystal lattice of $bcc$ type. The material and
structural parameter values used in the computations are as
follows: $M_{S,A}=0.8\; 10^{6}Am^{-1}$, $M_{S,B}=0.2\;
10^{6}Am^{-1}$, $A_{A}=0.04\; 10^{-11}Jm^{-1}$, $A_{B}=0.02\;
10^{-11}Jm^{-1}$, $f=0.25$, $a=50\AA$, $\mu_{0}H_{0}=0.1T$, close
to the respective estimate values obtained experimentally
\cite{[32]}. \label{fig7}}
\end{center} \end{figure}

We shall try to confront our theory with available experimental
results now. Energy gaps have recently been found in spin wave
spectra of doped manganites $La_{1-x}Ca_{x}MnO_{3}$ and
$La_{1-x}Sr_{x}MnO_{3}$ at low concentrations ($2\%<x<9\%$) of
$Ca$ and $Sr$ ions \cite{[32]}-\cite{[35]}. The quoted studies
demonstrated the existence of ferromagnetic nanoregions (referred
to as {\em droplets}) in these materials at temperatures below a
critical point; the space between the droplets is filled with a
different magnetic (ferromagnetic or antiferromagnetic) medium. We
suggest that the spin wave spectrum gap recently found
experimentally through neutron scattering on doped manganites
might be an indirect indication of periodic spatial distribution
of the ferromagnetic droplets. Assuming that the droplets
(material {\bf A}) are sphere-shaped and disposed in the nodes of
a {\em bcc} lattice, we have computed the magnonic spectrum in a
3D composite with structural and material parameter values close
to those reported in the experimental studies
\cite{[32]}-\cite{[35]}. The resulting spin wave spectrum is shown
in Fig. \ref{fig7}; the following parameter values are assumed:
$M_{S,A}=0.8 \; 10^{6}Am^{-1}$, $M_{S,B}=0.2\; 10^{6}Am^{-1}$,
$A_{A}=0.04\; 10^{-11}Jm^{-1}$, $A_{B}=0.02\; 10^{-11}Jm^{-1}$, $f
= 0.25$, $a=50\AA$, $\mu_{0}H_{0}=0.1$T. A narrow energy gap is
found in the obtained spectrum (between the first and the second
bands), in the same frequency range as the gap which was found in
the neutron scattering experiments \cite{[32]}-\cite{[35]}. The
gap resulting from our numerical computations lies between 1.27meV
and 1.49meV; the gap found experimentally had width 0.2meV and was
reported to lie in the vicinity of 1.65meV. As the "computed" gap
has width 0.22meV, the result obtained on the ground of our theory
can be deemed close to the experimental one. This preliminary
result fills us with optimism, as it allows us to propose a
working hypothesis supposing that the manganites in question can
be regarded as 3D manganite structures. This, however, will remain
unconfirmed as long as no direct experimental evidence of the
regularity of the structure formed by the droplets is available
({\em cf.} results reported in \cite{[36]}). We shall continue our
research in order to find further evidence supporting our
hypothesis.

\subsection*{Acknowledgements}

The present work was supported by the Polish State Committee for
Scientific Research through projects KBN - 2P03B 120 23 and
PBZ-KBN-044/P03-2001.


\end{document}